\documentclass[prl,tighten,article,twocolumn,showpacs]{revtex4}
\usepackage{amsmath,amsfonts,amssymb,times}
\usepackage{amsmath,amsfonts,amssymb}
\newcommand{\ketbrad}[1]{|#1\rangle\!\langle #1|}

\newcommand{\trace}{\mathop{\rm Tr}\nolimits}
\newcommand{\tr}{\mathop{\rm Tr}\nolimits}

\newcommand{\diag}{\mathop{\rm Diag}\nolimits}

\newcommand{\twomat}[4]{{\left(\begin{array}{cc}#1&#2\\#3&#4\end{array}\right)}}

\newcommand{\id}{\mathrm{\openone}}

\newcommand{\be}{\begin{equation}}
\newcommand{\ee}{\end{equation}}
\newcommand{\bea}{\begin{eqnarray}}
\newcommand{\eea}{\end{eqnarray}}
\newcommand{\beas}{\begin{eqnarray*}}
\newcommand{\eeas}{\end{eqnarray*}}

\newcommand{\qed}{\hfill$\square$\par\vskip24pt} % already defined

\newtheorem{theorem}{Theorem}
\newtheorem{lemma}{Lemma}

\begin{document}
\title{The Quantum Chernoff Bound}
\author{K.M.R.~Audenaert}
\affiliation{
Imperial College London, The Blackett Lab--QOLS,
Prince Consort Road, London SW7 2BW, U.K. \\
and Institute for Mathematical Sciences, Imperial College London, 53 Prince's Gate, London SW7 2PG, U.K.
\email{k.audenaert@imperial.ac.uk} }
\author{J.~Calsamiglia}
\affiliation{Grup de F\'{\i}sica Te\`{o}rica, Universitat Aut\`{o}noma de Barcelona, 08193 Bellaterra (Barcelona),
Spain}
\author{Ll.~Masanes}
\affiliation{Department of Applied Mathematics and Theoretical Physics, University of Cambridge, Wilberforce Road,
Cambridge CB3 0WA, U.K.}
\author{R.~Mu\~{n}oz-Tapia}
\affiliation{Grup de F\'{\i}sica Te\`{o}rica, Universitat Aut\`{o}noma de Barcelona, 08193 Bellaterra (Barcelona),
Spain}
\author{A.~Acin}
\affiliation{ICFO-Institut de Ciencies Fotoniques, Mediterranean Technology Park, 08860 Castelldefels (Barcelona),
Spain}
\author{E.~Bagan}
\affiliation{Grup de F\'{\i}sica Te\`{o}rica, Universitat Aut\`{o}noma de Barcelona, 08193 Bellaterra (Barcelona),
Spain}
\author{F.~Verstraete}
\affiliation{ Fakult\"at f\"ur Physik, Universit\"at Wien, Boltzmanngasse 5, 1090 Wien, Austria  }
\date{\today}

\begin{abstract}
We consider the problem of discriminating two different quantum states in the setting of asymptotically many
copies, and determine the optimal strategy that minimizes the total probability of error. This leads to the
identification of the quantum Chernoff bound, thereby solving a long standing open problem. The bound reduces to
the classical Chernoff bound when the quantum states under consideration commute. The quantum Chernoff bound is
the natural symmetric distance measure between quantum states because of its clear operational meaning and because
of the fact that it does not seem to share the undesirable features of other distance measures like the fidelity,
the trace norm and the relative entropy.
\end{abstract}
 \maketitle
%%%%%%%%%%%%%%%%%%%%%%%%%%%%%%%%%%%%%%%%%%%%%%%%%%%%%%%%%%%%%%%%%%%%%%%%%%%%%%%%%%%%%%%%%%%%%%%%%%%%%%%%%%%%%
%
\section{Introduction}
One of the most basic tasks in information theory is the discrimination of two different probability distributions:
given a source that outputs variables following one out of two possible probability distributions, determine which
one it is with the minimal possible error. In a seminal paper, Chernoff \cite{chernoff} solved this problem in the
asymptotic regime and showed that the probability of error $P_e$ in discriminating two probability distributions
decreases exponentially in the number of tests $n$ that one can perform: $P_e\sim \exp(-n\xi_{CB})$. The optimal
exponent $\xi_{CB}$ arising in the asymptotic limit is called the Chernoff bound \cite{fn}.
One of the virtues of the Chernoff bound
is that it yields a very natural distance measure between probability distributions; it is essentially the unique
distance measure in the ubiquitous situation of independent and identically-distributed (i.i.d.) random variables.

A quantum generalization of this result is highly desired. Indeed, the concept of randomness is much more
elementary in the field of quantum mechanics than in classical physics. Given the large amount of experimental effort
in the context of quantum information processing to prepare and measure quantum states, it is of fundamental
importance to have a theory that allows to discriminate different quantum states. Despite considerable effort,
this quantum generalization of the Chernoff bound has until now remained unsolved. The problem is to discriminate
two sources that output many identical copies of one out of two different quantum states $\rho$ and $\sigma$, and
the question is to identify the exponent arising asymptotically when performing the optimal test to discriminate
them. This task is so fundamental that it was probably the first problem ever considered in the field of quantum
information theory; it was solved in the one-copy case more than 30 years ago \cite{helstrom,holevo}. In this
paper, we finally identify the asymptotic error exponent when the optimal strategy for discriminating the states
is used. A nice feature of such a result is its universality, as it identifies the unique metric quantifying the
distance of quantum states in the i.i.d.\ setting \cite{footnote}.

Distance measures between quantum states have been used in a wide variety of applications in quantum information
theory. The most popular such measure seems to be Uhlmann's fidelity \cite{Uhlman}, which happens to coincide with the quantum
Chernoff bound when one of the states is pure. The trace distance has a more natural operational meaning, but
lacks monotonicity under taking tensor powers of its arguments. The problem is that one can easily find states
$\rho,\sigma,\rho',\sigma'$ such that ${\rm Tr}|\rho-\sigma|<{\rm Tr}|\rho'-\sigma'|$ but ${\rm Tr}|\rho^{\otimes
2}-\sigma^{\otimes 2}|>{\rm Tr}|\rho'^{\otimes 2}-\sigma'^{\otimes 2}|$. The quantum Chernoff bound exactly
characterizes the exponent arising in the asymptotic behaviour of the trace distance in the case of many identical
copies, and therefore does not suffer from this problem. Note that a similar situation happens in the case of
one-copy entanglement versus the asymptotic entanglement entropy.

In this work we give an upper bound for the probability of error for
discriminating two arbitrary states. In the particular case of a
large number of identical  copies, this result nicely complements
the recent work of Nussbaum and Szko{\l}a \cite{szkola}, where  a
lower bound for the asymptotic error exponent was found. These
respective upper and lower bounds coincide and hence give the exact
expression for the error exponent. The conjecture of Ogawa and
Hayashi concerning the quantum Chernoff bound raised in
\cite{hayashi} is thus solved.

Our paper is organized as follows. After the mathematical formulation of the problem, we prove a
nontrivial and fundamental inequality relating the trace distance to the quantum Chernoff bound. Finally, we prove
some interesting properties of the quantum Chernoff bound and discuss some applications.
%%%%%%%%%%%%%%%%%%%%%%%%%%%%%%%%%%%%%%%%%%%%%%%%%%%%%%%%%%%%%%%%%%%%%%%%%%%%%%%%%%%%%%%%%%%%%%
%
\section{Problem Formulation}
The optimal error probability of discriminating two quantum states $\rho_0$ and $\rho_1$ has been identified a long
time ago by Helstr\"om \cite{helstrom}.
We consider the two hypotheses $H_0$ and $H_1$ that a given quantum system is prepared either in the state $\rho_0$
or in the state $\rho_1$, respectively. Since the (quantum) Chernoff bound arises in a Bayesian setting, we
supply the prior probabilities $\pi_0$ and $\pi_1$, which are positive quantities summing up to 1
(the degenerate cases $\pi_0=0$ or $\pi_1=0$ are excluded).

Physically discriminating between these hypotheses corresponds to performing a generalised (POVM) measurement on the
quantum system with two outcomes, 0 and 1. This POVM consists of the two elements
$\{E_0,E_1\}$, where $E_0+E_1=\openone, E_i\geq 0$. The symmetric distinguishability problem consists in
finding those $E_0$ and $E_1$ that minimise
the total error probability $P_e$, which is given by
$$
P_e=\pi_0\trace[E_1\rho_0] + \pi_1\trace[E_0\rho_1].
$$

This problem can be solved using some basic linear algebra.
Let us first introduce some basic notations. Abusing terminology, we will use the term `positive' for
`positive semidefinite' (denoted $A\ge0$) in order to preserve trees. We employ the positive semidefinite ordering
throughout, $A\ge B$ iff $A-B\ge0$. The absolute value $|A|$ is defined as $|A|:=(A^* A)^{1/2}$. The Jordan
decomposition of a self-adjoint operator $A$ is given by $A=A_+ - A_-$, where $A_+$ and $A_-$ are the positive and
negative part of $A$, respectively, and are defined by. $A_+:=(|A|+A)/2$ and $A_-:=(|A|-A)/2$. Both parts are
positive by definition, and $A_+A_-=0$.

Note now that $P_e$ can be rewritten as
\begin{eqnarray*}
P_e &=& \pi_1 - \trace[E_1(\pi_1 \rho_1 - \pi_0 \rho_0)].
\end{eqnarray*}
This expression has to be minimised over all operators $E_1$ that satisfy $0\le E_1\le \id$.
The result is that $E_1$ has to be the projector on the range of the positive part of
$(\pi_1 \rho_1 - \pi_0 \rho_0)$.
We get
\beas
P_{e,\min}
&=& \pi_1 - \trace(\pi_1 \rho_1 - \pi_0 \rho_0)_+ \\
&=& \pi_1 - (\pi_1-\pi_0)/2 - \trace|\pi_1 \rho_1 - \pi_0 \rho_0|/2 \\
&=& \frac{1}{2}\left(1 - ||\pi_1 \rho_1 - \pi_0 \rho_0||_1 \right),
\eeas
where $||A||_1=\trace|A|$ is the trace norm.

The basic problem to be solved now is to identify how the error probability
$P_e$ behaves in the asymptotic limit, i.e.\ when one has to discriminate between
the hypotheses $H_0$ and $H_1$ corresponding to either $n$ copies of $\rho_0$ having been produced
or $n$ copies of $\rho_1$.
To do so, we need to study the quantity
$P_{e,\min,n}:=(1-||\pi_1 \rho_1^{\otimes n} - \pi_0 \rho_0^{\otimes n}||_1)/2$.

It turns out that the behaviour of $P_{e,\min,n}$ is exponential
\[
P_{e,\min,n}\sim \exp\left(-n \xi_{QCB} \right)
\]
and we will prove that the exponent $\xi_{QCB}$ is given by the following quantity, which can
therefore be called the \textit{quantum Chernoff bound}:
\begin{eqnarray}
\xi_{QCB}&=&\lim_{n\rightarrow\infty}-\frac{\log P_{e,\min,n}}{n} \label{distance} \\
&=&-\log\left(\min_{0\leq s\leq 1}\trace\left(\rho^{s}\sigma^{1-s}\right)\right).\label{lambda-chernoff}
\end{eqnarray}
Note that the quantity $\trace\left(\rho^{s}\sigma^{1-s}\right)$ is well defined and guaranteed to be positive.
As should be, this expression for the quantum Chernoff bound reduces to the usual definition of the classical
Chernoff bound $\xi_{CB}$ when $\rho$ and $\sigma$ commute: for classical distributions $p_0$ and $p_1$,
\be\label{chernoff-classic-1} \xi_{CB} = -\log\left(\min_{0\leq s\leq 1}\sum_i p_0(i)^{s} p_1(i)^{1-s}\right). \ee
It is truly remarkable that the quantum Chernoff bound is given by such a simple expression, looking like an
almost naive generalisation of the classical Chernoff bound with probabilities replaced by noncommuting quantum
states.

The fact that $\xi_{QCB}$ is lower bounded by the expression on the right hand side of
(\ref{lambda-chernoff}) was proven very
recently in \cite{szkola} (in a finite dimensional setting).
The fact that this is also an upper bound can be inferred from the following theorem, which
is the main contribution of this paper:

\begin{theorem}\label{th:1}
Let $A$ and $B$ be positive operators, then for all $0\le s\le 1$,
\be\label{eq:main}
\trace[A^s B^{1-s}] \ge \trace[A+B-|A-B|]/2.
\ee
\end{theorem}

Indeed, let $A=\pi_1\rho_1^{\otimes n}$ and $B=\pi_0\rho_0^{\otimes n}$, then the upper bound trivially follows
from the fact that the logarithm of the left hand side of the inequality (\ref{eq:main}) becomes
$\log(\pi_0^s\pi_1^{1-s})+n\log\left(\trace[\rho_0^s\rho_1^{1-s}]\right)$. Upon dividing by $n$ and taking the
limit $n\to\infty$, we obtain the quantum Chernoff bound $\xi_{QCB}$, independently of the priors $\pi_0$, $\pi_1$
(as long as the priors are not degenerate).

Inequality (\ref{eq:main}) is also very interesting from a purely matrix analytic point of view, as it relates the
trace norm to a multiplicative quantity that is highly nontrivial and very useful. Note that the optimal
measurement to discriminate the two sources enforces the use of joint measurements. The particular permutational
symmetry of $N$-copy states however guarantees that the optimal collective measurement can be implemented
efficiently (with a polynomial-size circuit) \cite{bacon04}, and hence that the minimum probability of error is
achievable with reasonable resources.

%%%%%%%%%%%%%%%%%%%%%%%%%%%%%%%%%%%%%%%%%%%%%%%%%%%%%%%%%%%%%%%%%%%%%%%%%%%%%%%%%%%%%%%
%
\section{Proof of Theorem 1}
Let now move on to prove Theorem \ref{th:1}. Note that the proof that we present here goes through in infinite
dimensions.

The proof relies on the following technical Lemma, which we prove in the Appendix A.
\begin{lemma}\label{lem:4}
Let $A,B\ge0$.
Let $0\le t\le 1$, and let $P$ be the projector on the range of $(A-B)_+$.
Then
\be\label{eq:lem4}
\trace[PB(A^t-B^t)]\ge0.
\ee
\end{lemma}

\textit{Proof of Theorem \ref{th:1}.} ---
We apply Lemma \ref{lem:4} to the case $t=s/(1-s)$, $A=a^{1-s}$ and $B=b^{1-s}$, where $a,b$ are positive operators
and $0\le s\le 1/2$.
With $P$ the projector on the range of $(a^{1-s}-b^{1-s})_+$, this yields
$$
\trace[Pb^{1-s}(a^s-b^s)]\ge0.
$$
Subtracting both sides from $\trace[P(a-b)]$ then yields
$$
\trace[a^s P (a^{1-s}-b^{1-s})]\le \trace[P(a-b)].
$$
Since $P$ is the projector on the range of the positive part of $(a^{1-s}-b^{1-s})$, the LHS can be rewritten as
$\trace[a^s (a^{1-s}-b^{1-s})_+]$. Because $a^s\ge0$, this is lower bounded by
$\trace[a^s (a^{1-s}-b^{1-s})] = \trace[a-a^s b^{1-s}]$.

On the other hand, the RHS is upper bounded by $\trace[(a-b)_+]$; this is because
for any self-adjoint $H$, $\trace[H_+]$ is the maximum of $\trace[QH]$ over all self-adjoint projectors $Q$.
We thus have
$$
\trace[a-a^s b^{1-s}] \le \trace[(a-b)_+] = \trace[(a-b)+|a-b|]/2.
$$
Subtracting both sides from $\trace[a]$ finally yields (\ref{eq:main}) for $0\le s\le 1/2$.
The remaining case $1/2\le s\le 1$ obviously follows by interchanging the roles of $a$ and $b$.
\qed

%%%%%%%%%%%%%%%%%%%%%%%%%%%%%%%%%%%%%%%%%%%%%%%%%%%%%%%%%%%%%%%%%
\section{Properties of the quantum Chernoff bound}

In this Section, we will also study the non-logarithmic variety of the quantum Chernoff bound, which we denote
here by $Q(\rho,\sigma):=\min_{0\le s\le 1} \trace[\rho^s \sigma^{1-s}]$.

We begin by stating upper and lower bounds on $Q$ in terms of the trace norm distance
$T(\rho,\sigma):=||\rho-\sigma||_1/2$; a proof can be found in the appendix B.
 \be 1-Q \le T\le
\sqrt{1-Q^2}. \ee Based on these bounds, the following properties of the $Q$-quantity and the Chernoff bound can
be derived:

\textit{Inverted measure.}
--- The maximum value $Q$ can attain is 1, and this is reached when $\rho=\sigma$. This follows, for example, from
the upper bound $Q^2+T^2\le 1$. The minimal value is 0, and this is only attained for pairs of orthogonal states,
i.e.\ states such that $\rho\sigma=0$. This implies that the Chernoff bound is infinite iff the states are
orthogonal; this has to be contrasted with the asymmetric error exponents occuring in the context of relative
entropy, where infinite values are obtained whenever the states have a different support.

\textit{Convexity in $s$.} --- The function to be minimised in $Q$ is $s\mapsto \trace[\rho^s \sigma^{1-s}]$. It
is important to realise that this function is convex in $s\in[0,1]$, because that means that the minimisation has
only one local minimum and therefore this local minimum is automatically the global minimum. This is an important
benefit in actual calculations.

Indeed, the function $s\mapsto x^s y^{1-s}$ is convex for positive scalars $x$ and $y$, as one easily confirms by calculating the second
derivative $x^s y^{1-s} (\log x-\log y)^2$, which is non-negative.
Consider then a basis in which $\rho$ is diagonal and given by $\rho=\diag(\lambda_1,\lambda_2,\ldots)$.
Let the eigenvalue decomposition of $\sigma$ (in that basis) be given by $\sigma=U \diag(\mu_1,\mu_2,\ldots) U^*$,
where $U$ is a unitary.
Then $\trace[\rho^s \sigma^{1-s}] = \sum_{i,j}\lambda_i^s \mu_j^{1-s} |U_{ij}|^2$.
As this is a sum with positive weights of convex terms $\lambda_i^s \mu_j^{1-s}$, the sum itself is also convex.

\textit{Joint concavity in $(\rho,\sigma)$.} --- By Lieb's theorem \cite{lieb}, $\trace[\rho^s \sigma^{1-s}]$ is
jointly concave in $(\rho,\sigma)$. Since the quantum Chernoff bound is the pointwise minimum of $\trace[\rho^s
\sigma^{1-s}]$ (over a fixed set, namely over $s\in[0,1]$), it is itself jointly concave as well. The Chernoff
bound is therefore jointly convex, just like the relative entropy.

\textit{Monotonicity under CPT maps.} --- From the joint concavity one easily derives the following monotonicity
property: for any completely positive trace preserving (CPT) map $\Phi$, \be Q(\Phi(\rho),\Phi(\sigma))\ge
Q(\rho,\sigma). \ee To prove this, one first notes that $Q$ is basis independent, i.e.\ is invariant under unitary
conjugations:
$$
Q(U\rho U^*,U\sigma U^*) = Q(\rho,\sigma).
$$
Secondly, $Q$ is invariant under addition of an ancilla: let $\tau$ be the (normalised) ancilla state, then
$$
Q(\rho\otimes\tau,\sigma\otimes\tau) = Q(\rho,\sigma);
$$
this is because
$\trace[(\rho\otimes\tau)^s(\sigma\otimes\tau)^{1-s}] = \trace[\rho^s \sigma^{1-s}]\trace[\tau]$.
Exploiting the Stinespring form of a CPT map, the monotonicity statement follows for general CPT maps
if we can prove it for the partial trace map.
As noted by Uhlmann \cite{uhlmann,carlenlieb},
the partial trace map can be written as a convex combination of certain unitary conjugations.
Monotonicity of $Q$ under the partial trace then follows directly from its concavity and its unitary invariance.

\textit{Continuity.} --- By the lower bound $Q+T\ge 1$, $1-Q$ is continuous in the sense that states that are
close in trace norm distance are also close in $1-Q$ distance: $0\le 1-Q\le T$.

\textit{Relation to Fidelity}
--- If one of the states is pure, then $Q$ equals the Uhlmann fidelity. Indeed, assume that
$\rho_1=|\psi\rangle\langle\psi|$ is pure, then the minimum of the expression ${\rm Tr}(\rho_1^s\rho_2^{1-s})$ is
obtained for $s=0$ and reduces to $\langle\psi|\rho_2|\psi\rangle$. As shown in the appendix B, the fidelity is
always an upper bound to $Q$.

\textit{Relation to the relative entropy} ---
Just as in the classical case, there is a nice connection between
the quantum relative entropy and the Chernoff bound. By differentiating the expression ${\rm
Tr}(\rho^s\sigma^{1-s})$ with relation to $s$, one observes that the minimum (which is unique due to convexity) is
obtained when
\[{\rm Tr}(\rho^s\sigma^{1-s}\log\rho)={\rm Tr}(\rho^s\sigma^{1-s}\log\sigma).\]
By the cyclicity of the trace, one easily verifies that this is equivalent to the condition that
\[S(\tau_s||\rho)=S(\tau_s||\sigma)\]
with $S(A||B)$ the quantum relative entropy ${\rm Tr}(A\log A-A\log B)$ and $\tau_s$ defined as
\begin{equation}
\tau_s=\frac{\rho^s\sigma^{1-s}}{{\rm Tr}\rho^s\sigma^{1-s}}.\label{interpo}
\end{equation}
Note that $\tau_s$ is not a state, because it is not even self-adjoint (except in the commuting case).
Nevertheless, as it is basically the product of two positive operators, it has positive spectrum, and its entropy and the relative entropies
used in (\ref{interpo})
are well-defined.
The value of $s$ for which both relative entropies coincide is the optimal value $s^*$.
This $\tau_{s^*}$ can be considered the quantum generalisation of the Hellinger arc and interpolates between two different quantum states,
albeit in a rather special (unphysical) way.

\textit{Metric.} The quantum Chernoff bound (or its non-logarithmic variety) between two  infinitesimally close states $\rho$ and $\rho-d\rho$
induces a metric that gives a geometrical structure to the state space.
In Appendix~C this metric is shown to be
$$
ds^2=1-\min_{0\leq s\leq 1}\tr[\rho^s(\rho-d\rho)^{1-s}]= \frac{1}{2}\sum_{ij} \frac{|\langle
i|d\rho|j\rangle|^2}{(\sqrt{\lambda_{i}}+\sqrt{\lambda_{j}})^2}
$$
where $\rho=\sum_i\lambda_i|i\rangle\langle i|$ is the eigenvalue decomposition of $\rho$.

%%%%%%%%%%%%%%%%%%%%%%%%%%%%%%%%%%%%%%%%%%%%%%%%%%%%%%%%%%%%%%%%%
\section{Conclusion}
We have identified the exact expression of the quantum generalization of
the Chernoff bound, which allows to quantify the asymptotic
behaviour of the error in the context of Bayesian discrimination of different
sources of quantum states. This resolves a long-standing open question.
Our main theorem (Theorem 1), which gives a computable lower bound to the trace norm difference of
two states in the many-copy regime,
may also find other relevant applications in and outside the field of
state discrimination.

\section{Acknowledgements}
FV and KA thank the hospitality of the Max Planck Institute for Quantum Optics where this work was initiated. KA
was supported by The Leverhulme Trust (grant F/07 058/U), by the QIP-IRC (www.qipirc.org) supported by EPSRC
(GR/S82176/0), by EU Integrated Project QAP, and by the Institute of Mathematical Physics, Imperial College
London. AA thanks the Spanish MEC, under a ``Ram\'on y Cajal" grant and the FIS2004-05639-C02-02 project, for
support. We are grateful to Montserrat Casas, Juli C\'espedes,  Alex Monr\`{a}s, Sandu Popescu and Andreas Winter
for discussions. We acknowledge financial support from the Spanish MCyT, under Ram\'on y Cajal program (AA and JC)
and FIS2005-01369, and CIRIT SGR-00185.

\appendix
\section{Appendix A}

\textit{Proof of Lemma \ref{lem:4}.} --- We exploit the integral representation
\begin{equation}
a^t = \frac{\sin(t\pi)}{\pi} \int_0^{+\infty} dx\,\, \frac{ax^{t-1}}{a+x},
\label{e1}
\end{equation}
which is valid for $a\ge0$ and $0\le t\le 1$ \cite{bhatia}. Extending this representation to positive operators in
the usual way, we get \beas
\trace[PB(A^t-B^t)] &=& \frac{\sin(t\pi)}{\pi} \int_0^{+\infty} dx\,\, x^{t-1} \\
&& \trace[PB(A(A+x)^{-1}-B(B+x)^{-1})]. \eeas If we can prove that the integrand is positive for all $x>0$, then
the integral itself is also positive. To do so, we first reduce the integrand to yet another integral and then
prove that the integrand of that integral is positive. Let $\Delta=A-B$. Now note \beas
\lefteqn{A(A+x)^{-1}-B(B+x)^{-1}} \\
&=& (B+\Delta)(B+\Delta+x)^{-1}-B(B+x)^{-1} \\
&=& \int_0^1 dt \,\,\frac{d}{dt}(B+t\Delta)(B+t\Delta+x)^{-1} \\
&=& \int_0^1 dt \,\,x(B+t\Delta+x)^{-1}\Delta(B+t\Delta+x)^{-1}. \eeas Here, the last equality can be shown as
follows: with $B':=B+t\Delta$,
%\beas
%\lefteqn{\frac{d}{dt}(B+t\Delta)(B+t\Delta+x)^{-1}} \\
%&=& \lim_{\epsilon\to 0} \frac{1}{\epsilon}[(B'+\epsilon\Delta)(B'+\epsilon\Delta+x)^{-1} - B'(B'+x)^{-1}] \\
%&=& \lim_{\epsilon\to 0} \frac{x}{\epsilon}[(B'+\epsilon\Delta+x)^{-1}(B'+\epsilon\Delta-B')(B'+x)^{-1}] \\
%&=& x(B'+x)^{-1}\Delta(B'+x)^{-1}. \eeas
%where we have used the easily checked identity
%$$
%A(A+x)^{-1}-B(B+x)^{-1} = x(A+x)^{-1}\Delta(B+x)^{-1}.
%$$
\beas
\lefteqn{\frac{d}{dt}(B+t\Delta)(B+t\Delta+x)^{-1}} \\
&=&\Delta(B'+x)^{-1} - B'(B'+x)^{-1}\Delta(B'+x)^{-1} \\
&=& x(B'+x)^{-1}\Delta(B'+x)^{-1}.
\eeas
Therefore, \beas
\lefteqn{\trace[PB(A(A+x)^{-1}-B(B+x)^{-1})]} \\
&=& x\,\int_0^1 dt \,\,\trace[PB(B+t\Delta+x)^{-1}\Delta(B+t\Delta+x)^{-1}], \eeas where we note that $B+t\Delta$
is positive for $0\le t\le 1$.

If we can show that the integrand is positive for $0\le t\le 1$, then the integral itself is also positive. In the
following we absorb $t$ in $\Delta$ and write $C$ as a shorthand for $B+t\Delta$. What we have to prove then is
that for all $B,C\ge0$, $x$ a non-negative scalar, and $P$ the projector on the range of $(C-B)_+$,
\be\label{eq:lem2} \trace[PB(C+x)^{-1}\Delta(C+x)^{-1}]\ge0. \ee Let $\Delta$ have the Jordan decomposition
$\Delta=\Delta_+ - \Delta_-$. Thus $P$ is the projector on the range of $\Delta_+$. We introduce the symbol
$V=(C+x)^{-1}\ge0$. We choose a basis in which $\Delta$ and $P$ can be partitioned as
$$
\Delta = \twomat{\Delta_+}{0}{0}{-\Delta_-}, \quad P=\twomat{\id}{0}{0}{0}.
$$
In that same basis, $V$ can be partitioned as
$$
V = \twomat{V_{11}}{V_{12}}{V_{21}}{V_{22}}.
$$
For ease of notation, the subscript $ij$ will henceforth refer to the $(i,j)$-th block of an operator valued
expression. Inequality (\ref{eq:lem2}) can then be rewritten as
$$
\trace[P(C-\Delta)(V\Delta V)] = \trace[(C-\Delta)(V\Delta V)]_{11} \ge0.
$$
Noting that $C-\Delta = V^{-1}-\Delta-x$, the LHS is equal to \bea
\lefteqn{\trace[(V^{-1}-\Delta-x)(V\Delta V)]_{11}} \nonumber \\
&=& \trace[\Delta V - (\Delta+x)(V\Delta V)]_{11} \nonumber \\
&=& \trace[\Delta_+ V_{11} - (\Delta_+ + x)(V\Delta V)_{11}] \nonumber \\
&=& \trace[\Delta_+ (V-V\Delta V)_{11}-x(V\Delta V)_{11}]. \label{eq:i1} \eea Because of the positivity of $B$, we
have $V^{-1}\ge \Delta+x$, which implies $V=VV^{-1}V\ge V(\Delta+x)V = V\Delta V+x V^2$. As the diagonal blocks of
a positive operator are themselves positive, this further implies
$$
V_{11} - (V\Delta V)_{11} \ge x (V^2)_{11}.
$$
Inserting this in (\ref{eq:i1}) gives \beas
\lefteqn{\trace[(V^{-1}-\Delta-x)(V\Delta V)]_{11}} \\
&=& \trace[\Delta_+ (V-V\Delta V)_{11}-x(V\Delta V)_{11}] \\
&\ge& \trace[\Delta_+ x (V^2)_{11}-x(V\Delta V)_{11}] \\
&=& x \trace[\Delta_+ (V^2)_{11} - (V\Delta V)_{11}] \\
&=& x \trace[\Delta_+ (V_{11}V_{11}+V_{12}V_{21}) - (V_{11}\Delta_+ V_{11} - V_{12}\Delta_- V_{21})] \\
&=& x \trace[\Delta_+ V_{12}V_{21} + V_{12}\Delta_- V_{21}]. \eeas By the fact that $V_{12}$ and $V_{21}$ are each
other's adjoint, the latter expression is positive, which finally proves the statement of the Lemma. \qed

\section{Appendix B}
Theorem \ref{th:1} applied to normalised states immediately gives the lower bound \be Q+T\ge 1. \ee Below we
provide an upper bound on $Q$ that is valid for any pair of states.

By definition, for any fixed value of $s$ between 0 and 1, the quantity $\trace[\rho^s \sigma^{1-s}]$ is an upper
bound on $Q$. In what follows, we set $s=1/2$. Furthermore, by replacing the trace with the trace norm, we get an
even higher upper bound. Indeed, \bea
Q&\le&\trace[\rho^{1/2}\sigma^{1/2}] \nonumber \\
&=& ||\rho^{1/4}\sigma^{1/2}\rho^{1/4}||_1 \nonumber \\
&\le& ||\rho^{1/2}\sigma^{1/2}||_1.\label{eq:QF} \eea In the last line we have used the fact (\cite{bhatia}, Prop.
IX.1.1) that for any unitarily invariant norm $|||AB|||\le |||BA|||$ if $AB$ is normal. In particular, consider
the trace norm, with $A=\rho^{1/4}\sigma^{1/2}$ and $B=\rho^{1/4}$; then $AB$ is self-adjoint, hence normal.

What we obtain as the RHS is the so-called Uhlmann fidelity $F$ between the states $\rho$ and $\sigma$:
$$
F:=\trace[(\rho^{1/2}\sigma\rho^{1/2})^{1/2}] = ||\rho^{1/2}\sigma^{1/2}||_1.
$$
This quantity can be regarded as the generalisation of the Bhattacharyya coefficient to the quantum case. For
classical distributions $p_0$ and $p_1$, the Bhattacharyya coefficient is defined as \cite{fuchs}
$B(p_0,p_1):=\sum_i \sqrt{p_0(i) p_1(i)}$. We have thus just shown that $F$ is an upper bound to $Q$.

Furthermore, by a result of Fuchs and van de Graaf \cite{fuchs}, $1-\sqrt{F}$ is lower bounded by the square of
the trace distance $T$: \be\label{eq:TF} T^2\le 1-\sqrt{F}. \ee Combining this with inequality (\ref{eq:QF})
yields the upper bound \be\label{eq:QT} Q^2+T^2\le 1. \ee

There is a nice direct proof of the latter inequality that circumvents the proof of (\ref{eq:TF}) and goes through
in infinite dimensions. We state it in terms of general positive operators:
\begin{theorem}
For positive operators $A$ and $B$, \be ||A-B||_1^2 + 4(\trace[A^{1/2} B^{1/2}])^2 \le (\trace(A+B))^2. \ee
\end{theorem}
\textit{Proof.} Consider two general operators $P$ and $Q$, and define their sum and difference as $S=P+Q$ and
$D=P-Q$. We thus have $P=(S+D)/2$ and $Q=(S-D)/2$. Consider the quantity \beas
PP^*-QQ^* &=& \frac{1}{4}\left((S+D)(S+D)^* - (S-D)(S-D)^*\right) \\
&=& \frac{1}{2}(SD^*+DS^*). \eeas Its trace norm is upper bounded as \beas
||SD^*+DS^*||_1/2 &\le& (||SD^*||_1 + ||DS^*||_1)/2 \\
&=& ||SD^*||_1 \\
&\le& ||S||_2 ||D||_2. \eeas In the last line we have used a specific instance of H\"older's inequality for the
trace norm (\cite{bhatia} Cor.\ IV.2.6). Now put $P=A^{1/2}$ and $Q=B^{1/2}$, which exist by positivity of $A$ and
$B$, and which are by themselves positive operators. We get $S,D=A^{1/2}\pm B^{1/2}$, hence
$$
||A-B||_1 \le ||A^{1/2} + B^{1/2}||_2 \,\,||A^{1/2} - B^{1/2}||_2,
$$
which upon squaring becomes \beas
||A-B||_1^2 &\le& \trace(A^{1/2} + B^{1/2})^2 \,\,\trace(A^{1/2} - B^{1/2})^2 \\
&=& \trace(A+B+A^{1/2}B^{1/2}+B^{1/2}A^{1/2})\\
&& \times \trace(A+B-A^{1/2}B^{1/2}-B^{1/2}A^{1/2}) \\
&=& (\trace(A+B)+2\trace(A^{1/2}B^{1/2}))\\
&& \times (\trace(A+B)-2\trace(A^{1/2}B^{1/2})) \\
&=& (\trace(A+B))^2 -4(\trace(A^{1/2}B^{1/2}))^2. \eeas \qed

Together with the lower bound $Q+T\ge 1$ we can now bracket the trace distance in function of the $Q$ quantity:
\be 1-Q \le T\le \sqrt{1-Q^2}\leq 1-Q^2/2, \ee where the last inequality  becomes a very good approximation for small values of $Q$.

\section{Appendix C}

\emph{Derivation of the quantum Chernoff metric:}
The goal is to calculate
\begin{equation}
ds^2=1-\min_{0\leq s\leq 1}\tr[\rho^s(\rho-d\rho)^{1-s}].
\label{ds2a}
\end{equation}

Here we will use the integral representation \eqref{e1} as well as  its derivative,
\begin{equation}
\label{e2}
t a^{t-1} = \frac{\sin(t\pi)}{\pi} \int_0^{+\infty} dx\,\, \frac{x^{t}}{(a+x)^2},
\end{equation}
which holds for $a\geq 0$ and $-1\leq t\leq 1$.

In particular, using \eqref{e1} and the convergent sequence
$$1/(a-b)=a^{-1} +a^{-1} b a^{-1}+a^{-1} b a^{-1} b a^{-1}+\ldots$$
one can write $(\rho-d\rho)^{1-s}$  up to second order in $d\rho$,
\begin{eqnarray}
(\rho-d\rho)^{1-s}&=& c_{s} \int_0^{+\infty} dx\,\, (\rho-d\rho) \frac{x^{-s}}{\rho-d\rho+x}\nonumber\\
&\approx& c_{s} \int_0^{+\infty} dx\,\, x^{-s} (\rho-d\rho)\left(\frac{1}{\rho+x}\right.\nonumber\\
&& \left.+\frac{1}{\rho+x}d\rho\frac{1}{\rho+x}+\frac{1}{\rho+x}d\rho\frac{1}{\rho+x}d\rho\frac{1}{\rho+x}\right),\nonumber
\end{eqnarray}
where $c_{s} =\pi^{-1}\sin[s\pi]$.

Inserting this expansion in \eqref{ds2a} one finds
\begin{eqnarray}
ds^2&=&\max_{0\leq s\leq 1} c_{s} \int_0^{+\infty} dx \tr\left[ \frac{x^{1-s}}{(\rho+x)^2}\rho^sd\rho\right.
\nonumber \\
&&\left.+ \frac{x^{1-s}}{(\rho+x)^2}\rho^sd\rho\frac{1}{\rho+x}d\rho\right].
\end{eqnarray}
The first term in the integrand vanishes, as can be seen by using \eqref{e2} and $\tr d\rho=0$,
while the second term can be computed in the basis that diagonalizes $\rho=\sum_{i}\lambda_{i}\ketbrad{i}$:
\begin{eqnarray}
ds^2&=& \max_{0\leq s\leq 1}  \sum_{ij} c_{s} \int_0^{+\infty} dx x^{1-s}
\frac{\lambda_{i}^s}{(\lambda_{i}+x)^2(\lambda_{j}+x)}|\langle i|d\rho|j\rangle
|^2\nonumber\\
&=&\max_{0\leq s\leq 1}  \sum_{ij} \frac{|\langle i|d\rho|j\rangle |^2}
{(\lambda_{i}-\lambda_{j})^2}(\lambda_{i}+\lambda_{j}-\lambda_{i}^{s}\lambda_{j}^{1-s}-\lambda_{j}^{s}\lambda_{i}^{1-s})\nonumber\\
&=&\frac{1}{2} \sum_{ij} \frac{|\langle i|d\rho|j\rangle
|^2}{(\lambda_{i}-\lambda_{j})^2}(\lambda_{i}+\lambda_{j}-2\sqrt{\lambda_{i}\lambda_{j}})\nonumber\\
&=&\frac{1}{2}\sum_{ij} \frac{|\langle i|d\rho|j\rangle |^2}{(\sqrt{\lambda_{i}}+\sqrt{\lambda_{j}})^2},
\end{eqnarray}
where in the second equality we have used $d\rho=d\rho^\dagger$ and  symmetrized the factor that multiplies
$|\langle i|d\rho|j\rangle |^2$ in the sum.

%%%%%%%%%%%%%%%%%%%%%%%%%%%%%%%%%%%%%%%%%%%%%%%%%%%%%%%%%%%%%%%%%

%------------------------------------------------------------- BIBLIOGRAPHY


\begin{thebibliography}{99}
\bibitem{fn}
The quantity which in this paper is referred to as the ``Chernoff bound'' also goes under the alternative names of
Chernoff distance, Chernoff divergence, and Chernoff information. While the term ``Chernoff bound'' is also used
for a bound on tail probabilities for statistical distributions (of which, actually, the Chernoff information is
a derived quantity) we decided, against common sense, to follow common terminology.
\bibitem{footnote}
A related problem that attracted a lot of attention in the field of quantum information theory was to identify the
quantum relative entropy between two quantum states. Classically, the relative entropy between two probability
distributions characterizes the inefficiency of compressing messages from a source $p$ with an algorithm that
would yield the Shannon information bound for a source $p'$. Phrased differently, it quantifies the way one could
cheat by telling that the given probability distribution is $p$ but the real one is $p'$. By proving a quantum
version of Stein's lemma \cite{ogawa}, it has been shown that the quantum relative entropy has exactly the same
operational meaning. One problem with this measure however is the fact that it is not continuous and is asymmetric
between its arguments, and does therefore not induce a natural distance measure between quantum states. The
quantum Chernoff bound seems to be much more natural in many situations.
\end{thebibliography}
\end{document}